\documentclass[]{pasj01}

\usepackage[left]{lineno}
\usepackage[T1]{fontenc}
\usepackage{bm}

\begin{document} 

\Received{2021/10/20}
\Accepted{2022/01/02}

\title{Structure of the super-Eddington outflow and its impact on the cosmological scale.}

\author{Ignacio \textsc{Botella}\altaffilmark{1}}
\email{ignacio@kusastro.kyoto-u.ac.jp}

\author{Shin \textsc{Mineshige}\altaffilmark{1}}
\author{Takaaki \textsc{Kitaki}\altaffilmark{1}}

\altaffiltext{1}{Department of astronomy, Kyoto University, Kitashirakawa-Oiwake-cho, Sakyo-ku, Kyoto-shi,Kyoto, 606-8502}

\author{Ken \textsc{Ohsuga},\altaffilmark{2}}
\altaffiltext{2}{Center for Computational Sciences, University of Tsukuba, 1-1-1 Tennodai, Tsukuba-shi,Ibaraki, 305-8577}

\author{Tomohisa \textsc{Kawashima}\altaffilmark{3}}
\altaffiltext{3}{Institute for Cosmic Ray Research, University of Tokyo, 5-1-5 Kashiwanoha, Kashiwa-shi,Chiba, 277-8582}



\maketitle

\begin{abstract}
It is one of the biggest issues in black hole (BH) astrophysics how to precisely evaluate BH feedback to its environments. Aiming at studying the unique gas dynamics of super-Eddington flow around supermassive black hole (SMBH) seeds at high redshift, we carried out axisymmetric two dimensional radiation hydrodynamic simulations by a nested simulation-box method. Here we divide the simulation box into the inner zone at $(2 - 3 \times 10^3) r_{\rm{Sch}}$ (with $r_{\rm Sch}$ being the Schwarzschild radius) and the outer zone at $(2\times 10^{3} - 3\times 10^6) r_{\rm{Sch}}$, with smooth connection of the physical quantities, such as gas density, velocity, and radiation energy. 
We start the calculation by injecting mass through the outer boundary of the inner zone at a constant rate of $\dot{M}_{\rm{inj}}=10^3L_{\rm{Edd}}/c^2$, where $L_{\rm{Edd}}$ is the Eddington luminosity and $c$ is the speed of light. A powerful outflow is generated in the innermost region and it propagates from the inner zone to the outer zone. The outflows are characterized by a velocity of 0.02$c$ (0.7$c$) and density of $10^{-17}$ ($10^{-19}$) g cm$^{-3}$ for near the edge-on (face-on) direction. The outflow is gradually accelerated as it travels by accepting radiation-pressure force. The final mass outflow rate at the outermost boundary is $\dot{M}_{\rm{out}}\sim 0.3 \times \dot{M}_{\rm{inj}}$. By extrapolating the outflow structure to a further larger scale, we find that the momentum and energy fluxes at $r \sim 0.1$ pc are  $\sim 10-100 L_{\rm{Edd}}/c $ and $\sim 0.1-10 L_{\rm{Edd}}$, respectively. Moreover, we find that the impacts are highly anisotropic in the sense that larger impacts are given towards the face-on direction than in the edge-on direction. These results indicate that the BH feedback will more efficiently work on the interstellar medium  than that assumed in the cosmological simulations.
\end{abstract}


\section{Introduction} \label{sec:intro}

We will start this section by introducing the objects of study, i.e., supermassive black hole (SMBH), and their associated problems. There are multiple observational studies conducted on SMBH across the universe (e.g., \cite{2010AJ....139..906W}, \cite{2016ApJS..227...11B}, \cite{2016ApJ...833..222J}, \cite{2017MNRAS.468.4702R}, \cite{2019ApJ...883..183M}, \cite{2019ApJ...871..199Y}, \cite{2019ApJ...884...30W}). Of these objects, there are some that present very large masses ($\sim 10^{9-10} M_\odot$) at a very early time ($z\sim6-7$) (\cite{2003AJ....125.1649F}, \cite{2018Natur.553..473B}). These objects, which can be studied to a certain extent through observation (\cite{2019arXiv190306106P}, \cite{2010A&ARv..18..279V}, \cite{2011BASI...39..145N}, \cite{2017PASA...34...31V}, and \cite{2020ARA&A..58...27I}), can not be satisfactorily answered through our current knowledge. Their massive size, so early in the universe, presents a challenge for our understanding of physics. These early objects are of particular importance since they defy our pre-existent knowledge of accretion physics or black hole formation. 

The SMBH in the early universe could form from the remnants of Population III (Pop III) stars, (\cite{1999ApJ...527L...5B}; \cite{2000ApJ...540...39A}; \cite{2001ApJ...548...19N}; \cite{2003ApJ...598...73Y}; \cite{2005MNRAS.363..379G}), or for the most massive seeds from direct gravitational collapse (\cite{2004MNRAS.354..292K}; \cite{2006MNRAS.370..289B}; \cite{2010Natur.466.1082M}; \cite{2011ApJ...742...13B}; \cite{2013ApJ...774..149C}). Other explanations can even have primordial haloes grow to masses
of $10^7 - 10^8 M_\odot$ and reach virial temperatures of $\sim 10^4 K$ without ever having formed a primordial (Pop III) star, either by being immersed in strong Lyman-Werner UV backgrounds that destroy all their $H_2$ (e.g., \cite{2012MNRAS.425.2854A}; \cite{2014MNRAS.440.1263Y}; \cite{2014MNRAS.442.2036D}; \cite{2014MNRAS.445..686J}; \cite{2015MNRAS.446.3163L}; \cite{2015MNRAS.454.2441S}, \cite{2017MNRAS.467.2288S}) or in highly supersonic baryon streaming motions that delay the collapse of the halo even if $H_2$ is present (\cite{2010PhRvD..82h3520T}; \cite{2011ApJ...736..147G}; \cite{2011ApJ...730L...1S}; \cite{2014MNRAS.443.1979L}; \cite{2017MNRAS.471.4878S}; \cite{2017Sci...357.1375H}).

But the problem common to all these studies is the lack of critical observational data. Without a reliable way to extract information of the physical situation around the high-z SMBH all conclusions can only be circumstantial. 

Thus, let us next take a look at what challenges observation of these objects pose. Observation of these SMBH can only be done through their hosts, the MeV blazars. These are the most luminous persistent sources in the Universe and emit most of their energy in the MeV band (\cite{2007ApJ...654..731H}, \cite{2013ApJ...770..103H}). These objects display very large jet powers and accretion luminosities. This points to the fact that relativistic jets may play a crucial role in rapid black hole growth at these early ages of the Universe (\cite{2019arXiv190306106P}).

Given that observation on these early objects are very limited the only tools available to study the physics involved are simulations. These simulations take 2 forms: large-scale cosmological simulations and small-scale astrophysical simulations. 

The first one of these tools that we need to introduce are the cosmological simulations. These simulations provide a very extensive study of the universe at any given redshift. In them the motion of gas and dark matter is calculated through hydrodynamical and N-body simulation respectively (eg. \cite{1997NewA....2..411P},  \cite{2000PhST...85...12T},\cite{2000MNRAS.317.1029P}, \cite{2019MNRAS.484.5437V}). This allows for the extensive study of structure formation on a very large range of scales ($\sim 1 \rm{Mpc}$ to $\sim 1 \rm{pc}$). The problem with these tools is their precision loss when studying smaller scale objects such as BH. This is undoubtedly due to the limitations on computational resources, where even the smaller cosmological simulations can not reach past $r\sim 0.1 \rm{pc}$ (\cite{2016MNRAS.456..500S}). It is well known that despite their small sizes BHs can produce large impacts on their environment in forms of radiation feedback and/or dynamical feedback by outflow. It is thus one of the biggest issues in BH astrophysics how to precisely evaluate such BH feedback on the cosmological scale.

Since following the evolution of large-scale outflow to smaller scales requires an enormous computational time, simulating until the boundary of the black hole is beyond impossible. Such a situation requires a simplification in which, to study the growth of the central object, all mass that passes the threshold (i.e., smallest resolution box) is considered to be absorbed. Other studies (\cite{2012MNRAS.420.2221D}) add a simplified outflow model as AGN feedback (non-relativistic, spherically symmetric wind), and find that the inclusion of this effect greatly impacts the medium they are in. Another issue with this approximation is the lack of outflow from the central object. As we already mentioned previously, SMBHs are known to present very bright relativistic jets. These objects can impact the structure of the surrounding gas (\cite{morabito2012jets}). For these reasons, cosmological simulations are not a competent tool to study these objects.  

The other tool, to analyze the evolution of SMBHs, that we need to introduce are astrophysical simulations. These simulations are focus on the study of accretion gas dynamics around the AGNs (e.g. \cite{1987ApJ...323..634E}, \cite{2015MNRAS.452.1502D}). These ones provide information of the infall and outflow structure though radio-hydrodynamic (RHD) physics. As a result, they create highly precise images of the gas interaction with the central object (e.g. \cite{1989A&A...222..141K}, \cite{1997PASJ...49..679O}, \cite{1998PASJ...50..639F}, \cite{1999ApJ...518..833K}, and \cite{2000PASJ...52L...5O}). Through them we can calculate the more precise rates of mass growth for the AGN as well as the strength of the outflow. Astrophysical simulations were done first in Newtonian dynamics (e.g. \cite{2009PASJ...61L...7O}, \cite{2011ApJ...736....2O}; \cite{2009PASJ...61..769K}; \cite{2014ApJ...796..106J}, \cite{2019ApJ...880...67J}) and later general relativistic treatment  was introduced (e.g. \cite{2014MNRAS.441.3177M}; \cite{2015MNRAS.447...49S}, \cite{2016MNRAS.456.3929S}; \cite{2016ApJ...826...23T}).

But similarly to the cosmological simulations, they also suffer from resource limitations. By being so precise, one can not use them to study the accretion physics far from the central object. This is due to the need to resolve the RHD equations (see Section 2) numerically. The larger the simulation box size is, and/or the smaller the resolution, the longer the simulation time becomes. In addition, to study the gas dynamics far enough from the central AGN, one needs to consider the change in the chemisty of the gas (e.g. \cite{2019MNRAS.488.2689T}; \cite{2020MNRAS.497..302T}), which adds extra complexity to the equations. 
Due to these factors mentioned above BH simulations, as their cosmological counterpart, are also rendered inadequate for tackling the SMBH growth question.

Previous work trying to bridge the gap between BH accretion and galactic simulations has been attempted before but with limitations. The simulations performed in \citet{2012ApJ...761..129Y} obtained a 2D mapping of hydrodynamics and MHD evolution of the gas. These simulations covered around 4 orders of magnitude in length scale by iterating the results between different-size simulation boxes. \citet{2020MNRAS.492.3272R} performed the 3D-general relativistic MHD simulations of Sgr A* by solving simulation from larger scale to smaller scale covering over 3 orders of magnitude, but they did not include the radiation impact. 

In this paper we will introduce a suitable methodology to link cosmological and astrophysical simulations to obtain a high precision long range RHD simulation. This methodology, what we call nested simulation-box method, will be developed and explained in Section \ref{sec:method}. Then in Section \ref{sec:results} we will show the results of applying this method to a scenario with a simplified accretion model and slow angular momentum infalling gas. In Section \ref{sec:discussion} we will discuss our results and findings. The final section will be devoted to conclusions.

\section{Methodology} \label{sec:method}

\subsection{Basic Equations of RHD} \label{subsec:RHD}

For this study we will use the simulation code developed in (\cite{2009PASJ...61..769K}, \cite{2005ApJ...628..368O} (Oshg+05)) as a basis to build our tool.  In this code a full set of axisymmetric two-dimensional RHD equations including the viscosity term are solved. The flux-limited diffusion (FLD) approximation is adopted (\cite{1981ApJ...248..321L}; \cite{2001ApJS..135...95T}). We also adopt the $\alpha$-viscosity prescription (\cite{1973A&A....24..337S}). General relativistic effects are incorporated by adopting the pseudo-Newtonian potential $\Psi(r) = -GM/(r - r_{\rm{Sch}})$ (\cite{1980A&A....88...23P}). Here, we assume that intermediate mass black holes are appropriate seeds of SMBHs, so that the mass of the central black hole is set to be $10^3 M_\odot$.

All equations are expressed in spherical polar coordinates. This coupled with other assumptions, non self-gravitating flow, reflection symmetric relative to the equatorial plane (with $\theta = \pi/2$), and axisymmetry with respect to the rotation axis (i.e., $\partial/\partial\varphi = 0$) creates a complete simplified set of main equations as following:

\noindent The continuity equation is

\begin{equation}
    \frac{\partial \rho}{\partial t}+\nabla\cdot(\rho\bm{v})=0.    
\end{equation}

\noindent The equations of motion are

\begin{equation}\label{r_motion}
   \frac{\partial (\rho v_r)}{\partial t}+\nabla\cdot(\rho v_r\bm{v})=-\frac{\partial p}{\partial r}+\rho \left(\frac{v_\theta^2}{r}+\frac{v_\phi^2}{r}-\frac{GM_{\rm{BH}}}{(r-r_{\rm{Sch}})^2} \right) \\
   + \frac{\chi}{c}F_{0,r},
\end{equation}

\begin{equation}
   \frac{\partial (\rho r v_\theta)}{\partial  t}+\nabla\cdot(\rho r v_\theta \bm{v})=-\frac{\partial p}{\partial \theta}+\rho v_\theta^2 \rm{cot}\theta\\
   +r\frac{\chi}{c}F_{0,\theta},
\end{equation}

\noindent and

\begin{equation}
  \frac{\partial (\rho r \sin \theta v_\phi)}{\partial  t}+\nabla\cdot(\rho r \sin \theta v_\phi \bm{v})=\frac{1}{r^2}\frac{\partial }{\partial r} \left( r^3 \rm{sin}\theta t_{r,\phi} \right),
\end{equation}

\noindent where $p$ is the gas pressure, $\chi = \kappa + \rho \sigma_T/m_{\rm{p}}$ is the total opacity, $\kappa$ is the free-free and bound-free absorption opacity (\cite{1986rpa..book.....R}), $\sigma_{\rm{T}}$ is the cross-section of Thomson scattering, $m_{\rm{p}}$ is the proton mass, and $F_0 = (F_{0,r}, F_{0,\theta}, F_{0,\phi})$ is the radiative flux in the comoving frame, where the suffix 0 represents quantities in the comoving frame. We set $F_{0,\phi} = 0$ because of the axisymmetry of the simulation.

We assume that only the $r$-$\phi$ component of the viscous-shear tensor is nonzero, and it is prescribed as

\begin{equation}
    t_{r,\phi}=\eta_r \frac{\partial}{\partial r}\left(\frac{v_\phi}{r}\right),
\end{equation}

\noindent with the dynamical viscous coefficient being:

\begin{equation}
    \eta_r=\rho\nu=\alpha\frac{p+\lambda E_0}{\Omega_{\rm{K}}}.
\end{equation}
    
\noindent Here, $\alpha = 0.1$ is the $\alpha$ parameter (\cite{1973A&A....24..337S}), $\Omega_K$ is the Keplerian angular speed, $E_0$ is the radiation energy density, and $\lambda$ represents the flux limiter of the flux-limited diffusion approximation (\cite{1981ApJ...248..321L}; \cite{2001ApJS..135...95T}):

The energy equation of the gas is

\begin{equation}
    \frac{\partial e}{\partial t}+\nabla\cdot(e\bm{v})=-p\nabla\cdot\bm{v}-4\pi\kappa B+c\kappa E_0\\
    +\Phi_{\rm{vis}}-\Gamma_{\rm{Comp}},
\end{equation}

\noindent and the energy equation of the radiation is

\begin{equation}
    \frac{\partial E_0}{\partial t}+\nabla\cdot(E_0\bm{v})=-\nabla\cdot\mathbf{F}_0-\nabla \bm{v}:\mathbf{P}_0+4\pi\kappa B-c\kappa E_0\\
    +\Gamma_{\rm{Comp}}.
\end{equation}

\noindent Here, $e$ is the internal energy density which is linked to the thermal pressure by the ideal gas equation of state, $p = (\gamma - 1)e = \rho k_{\rm{B}} T_{\rm{gas}}/(\mu m_{\rm{p}})$ with $\gamma = 5/3$ being the specific heat ratio, $k_{\rm{B}}$ the Boltzmann constant, $\mu = 0.5$ is the mean molecular weight (we assume pure hydrogen plasma), and $T_{\rm{gas}}$ is the gas temperature. $B = \sigma_{\rm{SB}} T^4_{\rm{gas}}/\pi$ is the blackbody intensity where $\sigma_{\rm{SB}}$ is the Stefan–Boltzmann constant, $P_0$ is the radiation pressure tensor, $\Phi_{\rm{vis}}$ is the viscous dissipative function written as:

\begin{equation}
    \Phi_{\rm{vis}}=\eta_r\left[ r \frac{\partial}{\partial r} \left( \frac{v_\phi}{r} \right) \right]^2.
\end{equation}

\noindent The Compton cooling/heating rate $\Gamma_{\rm{Comp}}$ is described as 

\begin{equation}
    \Gamma_{\rm{Comp}}=4\sigma_{T}c\frac{k_{\rm{B}}(T_{\rm{gas}}-T_{\rm{rad}})}{m_{\rm{e}}c^2} \left( \frac{\rho}{m_{\rm{p}}} \right) E_0.
\end{equation}

\noindent For these equations, $m_{\rm{e}}$ is the electron mass and $T_{\rm{rad}} \equiv (E_0/a)^{1/4}$ is the radiation temperature with the radiation constant $a = 4\sigma_{\rm{SB}}/c$. 

\subsection{Nested simulation-box method}\label{subsec:nested}

In this subsection, we explain our nested structured simulation box, which is introduced to bridge the gap between the astrophysical and cosmological simulations.

\subsubsection{Basic idea}

In order to quantitatively address the effects of the mechanical and radiative feedback on cosmological scale fluid, the outer boundary should be placed at $r \sim 0.1 \rm{pc} \sim 10^{9}$ $r_{\rm{Sch}} (M_{\rm{BH}}/10^3 M_\odot)^{-1}$. We, however, set the outer boundary at $r=10^6 r_{\rm{Sch}}$, because it is far enough from the black hole to evaluate the feedback effects as will be discussed at the end of Section \ref{sec:results}. The simulation box still covers an enormous length scale, so we need to implement a suitable method  by extending our RHD code. This method consists of a series of simulation boxes tied together to give a sense of zoom-out. The information is passed from one simulation box to the next one through boundary conditions. This allows us to trace the evolution of the inflow and outflow at different scales with maximum resolution and minimum computational cost.

\begin{figure*}[ht!]
    \begin{centering}
        \includegraphics[width=1\textwidth]{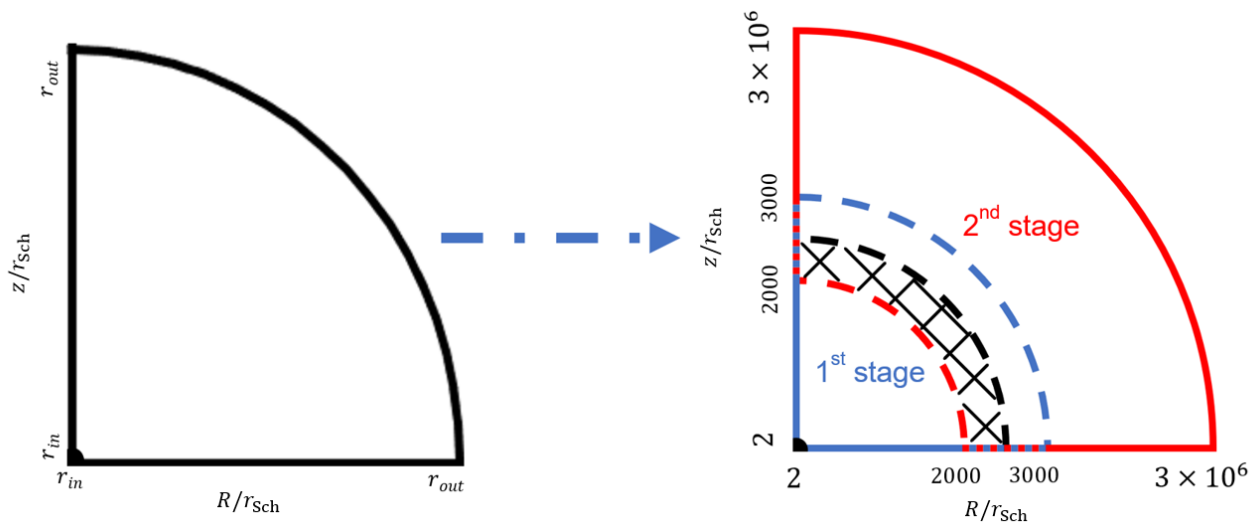}
    \end{centering}
    \caption{Schematic example of the nested simulation box method. The initial box (left) is separated into 2 smaller boxes (right), i.e., 1$^{st}$ stage (blue delimited area) ($2<R/r_{\rm{Sch}}<3000$) and 2$^{nd}$ stage (red delimited area) ($2000<R/r_{\rm{Sch}}<3\times 10^6$). The black marked area represents the connection zone shared between 1$^{st}$ to 2$^{nd}$ stage as boundary condition ($2000<R/r_{\rm{Sch}}<2500$). The remainder of the shared space will be used to determine the consistency between stages.}
    \label{AID_v2}
\end{figure*}

We prepare two zones: inner and outer zones (Table \ref{table1} and Figure \ref{AID_v2}), and perform simulations in two stages. 

\begin{table}[ht!]
\centering
\begin{tabular}{ |c|c|c|c|c|}
 \hline
 Stage & Zone & $r_{in}$ [$r_{\rm{Sch}}$] & $r_{out}$ [$r_{\rm{Sch}}$] & Simulated flow \\ 
 \hline
 1$^{st}$ & Inner &  $2$ & $3\times 10^3$ & inflow \& outflow\\  
 \hline
 2$^{nd}$ & Outer &  $2\times 10^3$ & $3 \times 10^6$ & outflow\\
 \hline
\end{tabular}
 \caption{Naming convention for the stages in the nested box method for the simulation discussed in this paper.}
 \label{table1}
\end{table}

The boundary conditions of these 2 simulation boxes are as follows: 

\begin{itemize}
    \item 1$^{st}$ stage inner boundary: Absorbing inner boundary (see Ohsg+05).
    \item 1$^{st}$ stage outer boundary: Free boundary (i.e., gas can escape the box), with inflow input through the equatorial plane as explained in section \ref{subsec:initial} (see also Ohsuga et al. 2005). 
    \item 2$^{nd}$ stage inner boundary: Connection zone to incorporate the results of 1$^{st}$ simulation (see subsection \ref{subsec:connection})
    \item 2$^{nd}$ stage outer boundary: Free boundary (i.e., gas can escape the box).
\end{itemize}

\subsubsection{Connection zone}\label{subsec:connection}

Here, we explain how to make a smooth connection between the two stages. The first step in this connection process is to take a set of time dependent variables of the gas and radiation (i.e., density, velocity vector, gas and radiation energy) at the moment when the 1$^{st}$ stage simulation has achieved quasi-steady state ($t_{\rm{qs}}$). We then time-average the physical variables ($\rho,\bm{v},E_{\rm{gas}}, E_{0}$) for $2000 r_{\rm{Sch}} \leq r \leq 2500 r_{\rm{Sch}}$ covering all $\theta$. This is to ensure a smooth connection between stages. 

Since we are only concerned with outflow properties to study AGN feedback, we do not consider inflow gas in the 2$^{nd}$ simulation. For this we re-scale the azimuthal angle ($\theta$) for $\theta > \theta_{\rm{ah}}$, with $\theta_{\rm{ah}}$ defined as the angular height of the inflow-outflow interface in the connection zone defined by $\theta | v_r(\theta)=0$. We set a gradual remapping of the values in the connection zone such that at the inner boundary of the 2$^{nd}$ stage (at $r = r_1 = 2000 r_{\rm{Sch}}$):

\begin{equation}
    x^{2\rm{nd}}(r_1,\theta)=x^{1\rm{st}}(r_1,\theta) \; \mathrm{for} \; 0<\theta<\pi/2, 
\end{equation}

\noindent but at the outer boundary of the connection zone (at $r = r_n = 2500 r_{\rm{Sch}}$):

\begin{equation}
    x^{2\rm{nd}}(r_n, \theta*)=x^{1\rm{st}}(r_n,\theta) \; \mathrm{for} \; 0<\theta<\pi/2 \\
    \mathrm{where} \; \theta*=\pi/2 \left( \frac{\theta}{\theta_{\rm{ah}}} \right).
\end{equation}

\noindent where $x$ represents the time-averaged hydrodynamic variables ($\rho,\bm{v},E_{\rm{gas}}$) and the super-index indicates the stage. This eliminates the inflow area by "stretching" the surrounding outflow values thus effectively remapping the inflow to grid points outside the simulation box.

The same method is employed for the radiation energy. The inner boundary condition in the 2$^{nd}$ simulation is set so as to also recreate the radiation profile seen in the previous stage (i.e., we no longer adopt the absorbing boundary used in the 1st simulation). The radiation flux is then solved from radiation energy through the FLD method.

We also need to consider that, the farther away from the black hole one places the gas, the colder it will be. This basic radiation trend means the gas may cool down beyond the $T=10^4\rm{K}$. This means that the approximation taken previously of the gas being fully ionized no longer stands true. This introduces new dispersion factors with the introduction of b-f interactions. In order to introduce this new factor to the code we added a modifier to the opacity factor:

\begin{equation}
    \kappa'=\kappa \left(\frac{\tanh\left(\frac{T_{\rm{gas}}-T_{\rm{ion}}}{\tau}\right)+1}{2}\right),
\end{equation}

\noindent where $T_{\rm{ion}}=10^4\rm{K}$, and $\tau=2\times10^3\rm{K}$ is a transition factor between the f-f and the b-f phase. This is necessary since when $T<10^4$K the hydrogen is not-ionized.
Therefore, the amount of the free electrons is smaller in this scenario, and the effective scattering opacity should become small.

Note that we only trace the outflow until the radius $= 3 \times 10^6 r_{\rm Sch}$, where all gas is free from the influence of the BH. This way we can then extrapolate that the structure will follow the same radial trends from that point on (as will be discussed at the end of Section \ref{sec:results}). 

\subsection{Initial setup}\label{subsec:initial}

We assume an atmosphere that is hot, optically thin, isothermal and in hydrostatic equilibrium in the radial direction, with negligible mass around the black hole (Ohsg+2005). The coronal gas is set in such a way that its atmospheric pressure does not impede gas flow, thus we start the simulation with $T_{\rm{gas}} = 10^{11}$ K and  $\rho = 10^{-17} \rm{g}\;\rm{cm}^{-3}$.

In the 1st stage simulation, the matter is injected from the outer boundary near the equatorial plane with $1.5<\theta\leq \pi/2$.  The injected matter has the low angular momentum corresponding to the Keplerian radius of $r_{\rm{Kep}} = 100 r_{\rm{Sch}}$ and super-Eddington mass accretion rate of $\dot{M}_{\rm{inj}} = 10^3 L_{\rm{Edd}}/c^2$, where $r_{\rm{Kep}}$ is the Keplerian radius, at which the centrifugal force (with a given specific angular momentum) is balanced with the central gravity.

In the 2$^{nd}$ stage simulation, the initial conditions of the coronal gas are taken as a thinner and much colder gas ($\rho = 10^{-21} {\rm{g}\; \rm{cm}^{-3}}$ and $T_{\rm{gas}} = 10^{4}$ K). The mass and radiation are injected from the connection zone near the inner boundary as is described in the section \ref{subsec:connection}.

\section{Results} \label{sec:results}

In this section we will present the results from the simulation performed with the nested simulation-box method. The results will be presented starting with the inflow part and then the outflow part. For the following results we will use the naming convention established in Table \ref{table1}. To save computation time, we assume symmetry of the flow structure with respect to the equatorial plane; namely, we have chosen $\theta\in\{0-\pi/2\}$ , in which $\theta = 0$ (or $\pi/2$) corresponds to the rotation axis (the equatorial plane). 

\subsection{Inflow properties}\label{subsec:1st_inflow}

In the first stage, the injected gas from the outer boundary rapidly falls unimpeded in a free-fall. This will continue all the way down to the region around the Keplerian radius, $r_{\rm Kep} = 100 r_{\rm Sch}$). Then, viscous (slow) accretion process starts. Because of this free-fall nature down to the Keplarian radius, the gas can be initiated at the 1$^{st}$ stage without altering the results. 

\begin{figure}[ht!] 
  \begin{centering}
    \includegraphics[width=0.5\textwidth]{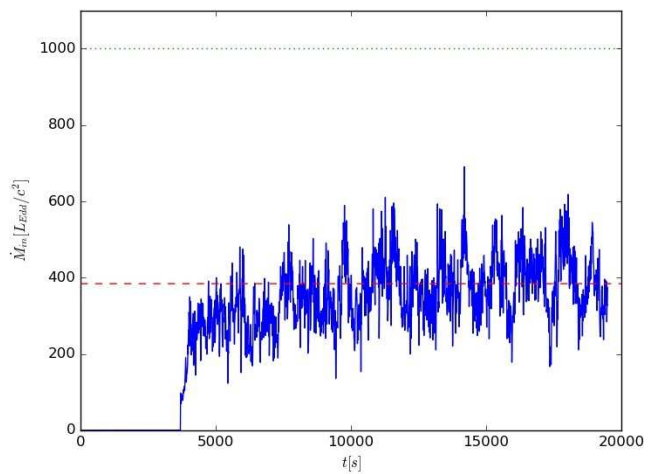}
  \end{centering}
    \caption{Time variations of the mass inflow rate onto the black hole ($\dot{M}_{\rm{BH}}$). The blue line traces the evolution of the inflow matter at the black hole boundary ($r=2r_{\rm{Sch}}$). The red dashed line represents the mean value of the blue line but in the quasi-steady state. The green dotted line indicates the mass injection rate at the outer boundary of the simulation box in the inner zone ($r=3 \times 10^3 r_{\rm{Sch}}$).}
    \label{M_in}
\end{figure}

From Figure \ref{M_in} (which shows time variations of the mass accretion rate onto the BH) we can see how the BH mass grows over time. While the mass flux shows fluctuations over time, we see that its time average settles down in a constant value at later times, indicating that a quasi-steady state is achieved. 
Since the flow must be in the quasi-steady state before proceeding to the next stage simulation, we need to confirm if it is indeed the case.
This can be done by looking at the mass flux profiles calculated as follows:

\begin{equation}
    \dot{M}_{\rm{in}}(r)\equiv 4\pi\int^{\pi/2}_0 \sin\theta r^2\rho(r,\theta) {\rm min}\{v_r(r,\theta),0\} d\theta,
\end{equation}

\begin{equation}
    \dot{M}_{\rm{out}}(r)\equiv 4\pi\int^{\pi/2}_0 \sin\theta r^2\rho(r,\theta) {\rm max}\{v_r(r,\theta),0\} d\theta,
\label{Mout_eq}
\end{equation}

\begin{equation}
    \dot{M}_{\rm{esc}}(r)\equiv 4\pi\int^{\pi/2}_0 \sin\theta r^2\rho(r,\theta) v_r' d\theta,
\end{equation}

\begin{equation}
    \dot{M}_{\rm{net}}(r) \equiv \dot{M}_{\rm{in}}+\dot{M}_{\rm{out}},
\end{equation}

\noindent where $\dot{M}_{\rm{in}}$, $\dot{M}_{\rm{out}}$, $\dot{M}_{\rm{esc}}$ and $\dot{M}_{\rm{net}}$ are the inflow rate, the outflow rate, the escape rate, and the net accretion rate, respectively, and $v_r'$ is defined as $v_r'=v_r$ for $v_r \geq v_{\rm{esc}}$ and $v_r' =0$ otherwise. (Note that the BH mass accretion rate corresponds to
$\dot{M}_{\rm{BH}}(t) \equiv |\dot{M}_{\rm{in}}(r=2r_{\rm{Sch}}, t)|$).

\begin{figure}[ht!]
  \begin{centering}
    \includegraphics[width=0.5\textwidth]{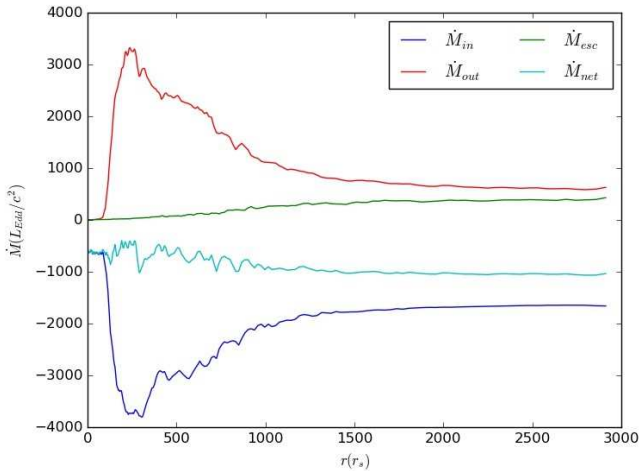}
  \end{centering}
    \caption{Radial trend of the mass inflow/outflow rate in the 1$^{st}$ stage simulation. The mass inflow rate ($v(r,\theta)<0$ ), the mass outflow rate ($v(r,\theta)>0$), the escaping mass outflow rate ($v(r,\theta)\geq v_{\rm{esc}}$) and the net mass flow rate are indicated by the dark blue, green, red, and cyan lines, respectively.}
    \label{M_dot_1st}
\end{figure}

In order to plot the integrated mass flux we average the integral over time (instead of integrating the time averages), in accordance to \citet{1999MNRAS.310.1002S}. For the quasi-steady state to be completely achieved in a simulation we should see that ${\dot M}_{\rm net} \simeq \rm{cnst}$, however this is not realistic (since it would require more than a year of computational time) nor completely necessary. We see, in Figure \ref{M_dot_1st}, that the net rate presents 2 distinct flow patterns, inside and outside the Keplerian radius. First we show how mass flows at a rate of $\dot{M}\sim 1000 (L_{\rm Edd}/c^2)$, which then gets divided in the two flows: inflow ($\sim 400 (L_{\rm Edd}/c^2)$), and outflow ($\sim 500 (L_{\rm Edd}/c^2)$). While the total simulation has not reached the quasi-steady state (i.e., circulating flow region net flux is not constant), the inflow and outflow regions have ($\dot{M}_{\rm{net}}(r<100)\sim \rm{cnt}$ \&  $\dot{M}_{\rm{net}}(r>1000)\sim \rm{cnt}$). 

In Figure \ref{M_dot_1st} we can also observe that the mass accepted by the BH is only a fraction of the injected mass per time. The remaining part will be circulated or ejected as outflow (see below).

\begin{figure*}[htbp!]
  \begin{centering}
    \includegraphics[width=1\textwidth]{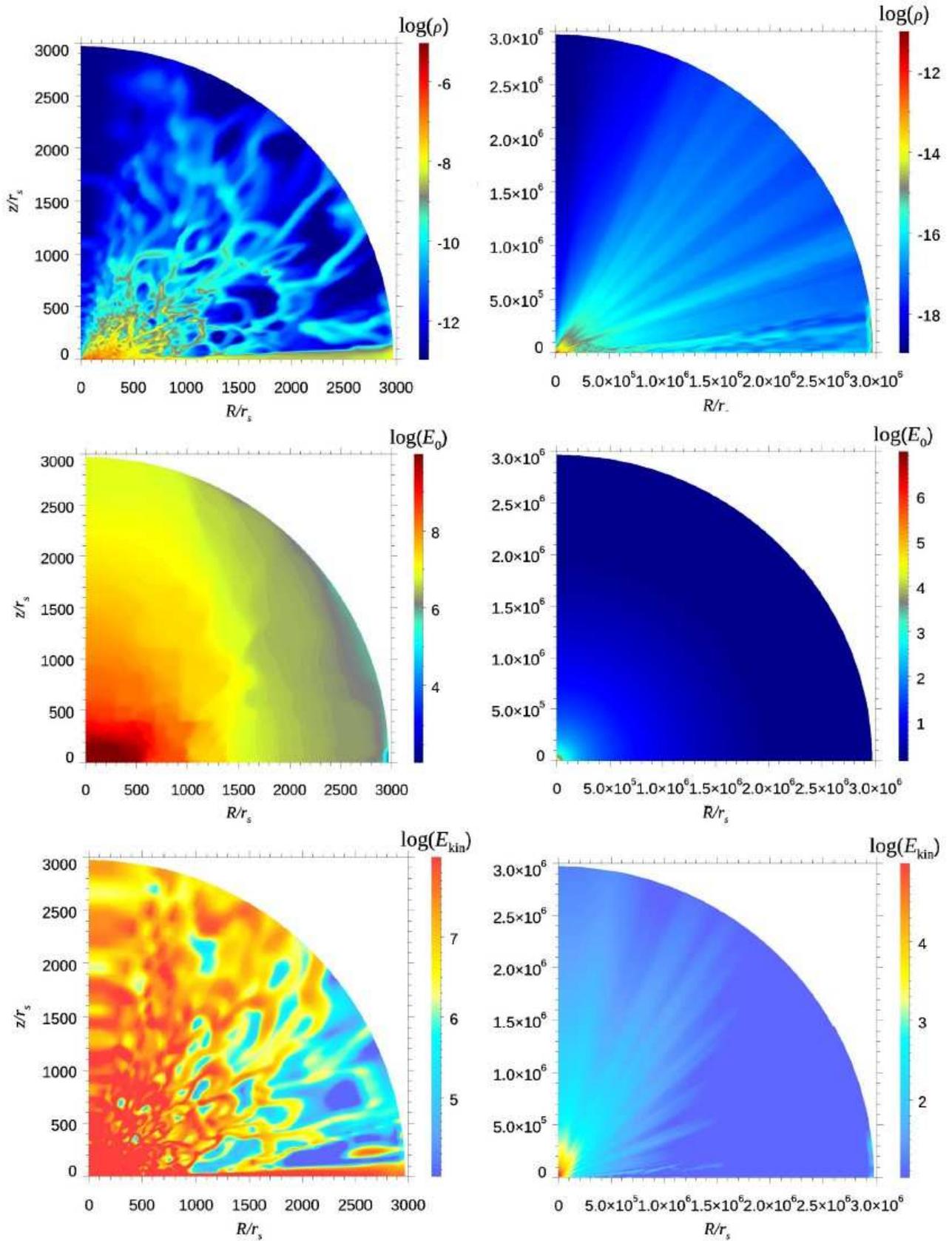}
  \end{centering}
  \caption{Two-dimensional (2D) contours of matter density (top), radiation energy density (middle), and kinetic energy density (bottom) in the inner zone (left) and outer zone (right). Note different color bars for the left and right panels; they are adjusted to clearly visualize rapid spatial variations of the physical quantities.}
    \label{outflow_steps}
\end{figure*}
 
\begin{figure}[ht!]
  \begin{centering}
    \includegraphics[width=0.48\textwidth]{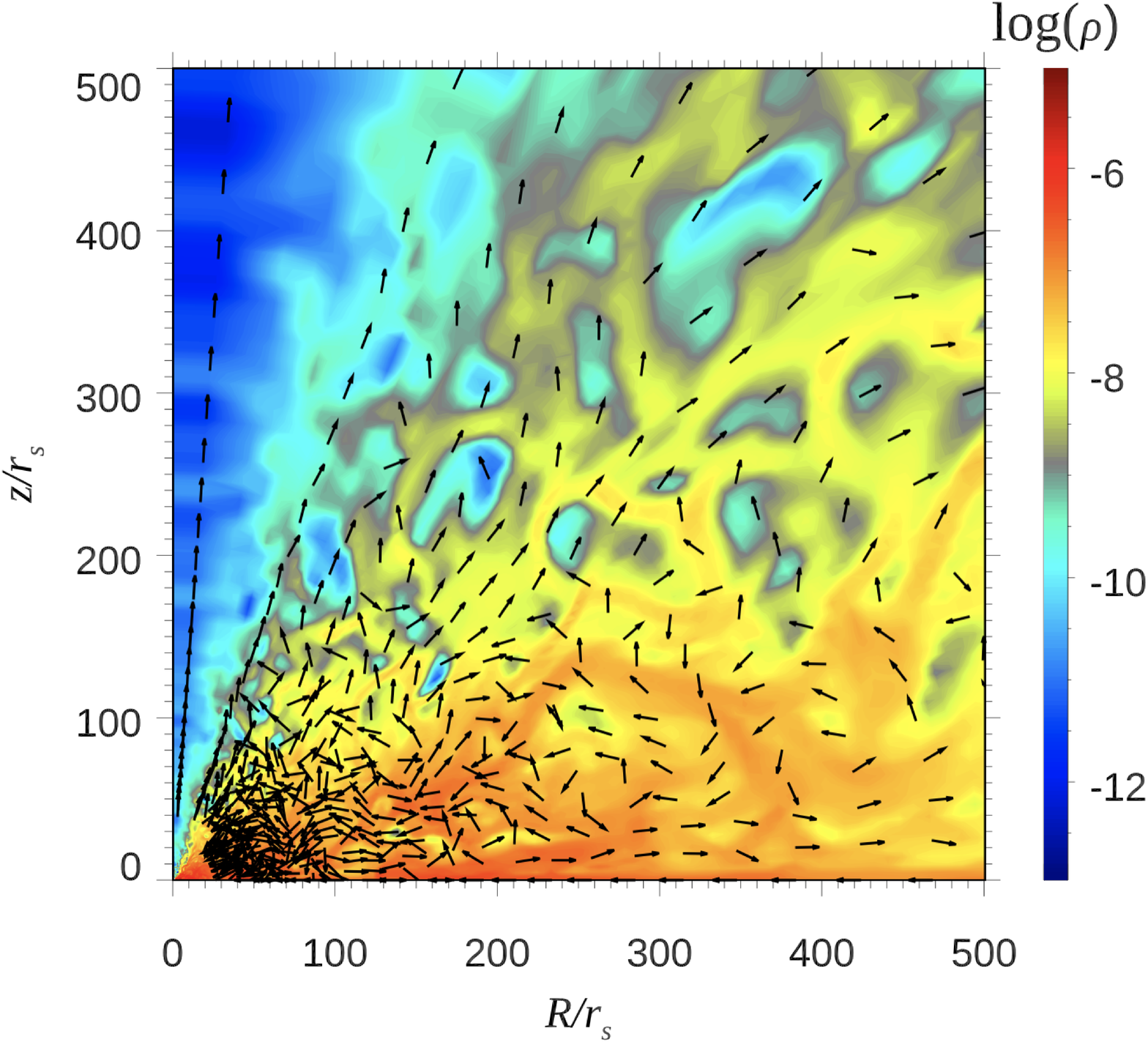}
  \end{centering}
  \caption{A typical snapshot of the two-dimensional (2D) contours of matter density in the 1$^{st}$ stage (at $t=18000 \, s$) centered around the {\lq}bulge{\rq}. The black arrows show the velocity vector map of each grid point.}
    \label{bulge}
\end{figure}

Time averaged inflow-outflow structure in the 1$^{st}$ stage simulation is summarized in the left panels of Figure \ref{outflow_steps}: from top to bottom, two-dimensional (2D) distributions of the gas density, the radiation energy density ($E_0$), and the kinematic energy, respectively. In the upper left panel we see a {\lq}bulge{\rq} (or puffed-up) structure, which is created by a small-scale circulation of gas formed between $100r_{\rm{Sch}}<r<1000r_{\rm{Sch}}$ (see Figure \ref{bulge}). 
It is important to note that such an inflated structure is created when the Keplerian radius is relatively small, as was demonstrated by \citet{2021arXiv210111028K}. We also observe low-density atmosphere surrounding the bulge structure, although it is not a static atmosphere but is composed of outflow (explained later).

\subsection{Outflow properties in the 1$^{st}$ stage}

Once the gas flow has reached the innermost region, outflow processes begin to produce powerful winds. In this section, we will trace the outflow in the 1$^{st}$ stage simulation.

There exists circulating flow patterns extending close to the outer boundary in the 1$^{st}$ stage simulation, although it is not always visible in the time-averaged plots of Figure \ref{outflow_steps} but we can confirm its presence in Figure \ref{M_dot_1st}, since a large bump in the ${\dot M}_{\rm{out}}$ curve and a large hollow in the ${\dot M}_{\rm{in}}$ curve are formed by the large-scale circulation flow. It is thus important to note how to choose the outer radius of the 1$^{st}$ stage simulation. If the radius of the outer boundary is chosen to be smaller than the circulating zone, mass will flow out of the simulation box and be lost, leading to an underestimation of $\dot{M}_{\rm{BH}}$. While a bigger simulation box could contain the larger scale phenomena, we would then be obliged to adopt coarse grid-point spacing to perform simulations within a reasonable time.
This would then result in missing details in the flow structure. Considering these facts, we have fixed the outer boundary of the 1st simulation to be at $3000 r_{\rm{Sch}}$.

In the middle left panel of Figure \ref{outflow_steps} we see that the radiation energy density monotonically decreases outward in a nearly spherically symmetric fashion. More precisely, the constant $E_0$ contours show a bit elongated in the vertical direction, which indicates radiation is going out more dominantly in the vertical ($z$) direction. The kinetic energy distributions displayed in the lower left panel, by contrast, show somewhat distinct patterns. First of all, the inflow (disk) region is clearly visualized, since not the radial velocity but the rotational ($v_\phi$) velocity is dominant and is comparable to the free-fall velocity. Second, outflow region, in which large $E_{\rm{kin}}$ is found, is rather elongated in the vertical direction. Third, the kinetic energy is at minimum above the disk region at large radii. To summarize, kinetic energy is released predominantly in the perpendicular direction to the disk plane. This feature will further be examined in the next subsection.

\begin{figure}[ht!] 
  \begin{centering}
    \includegraphics[width=0.45\textwidth]{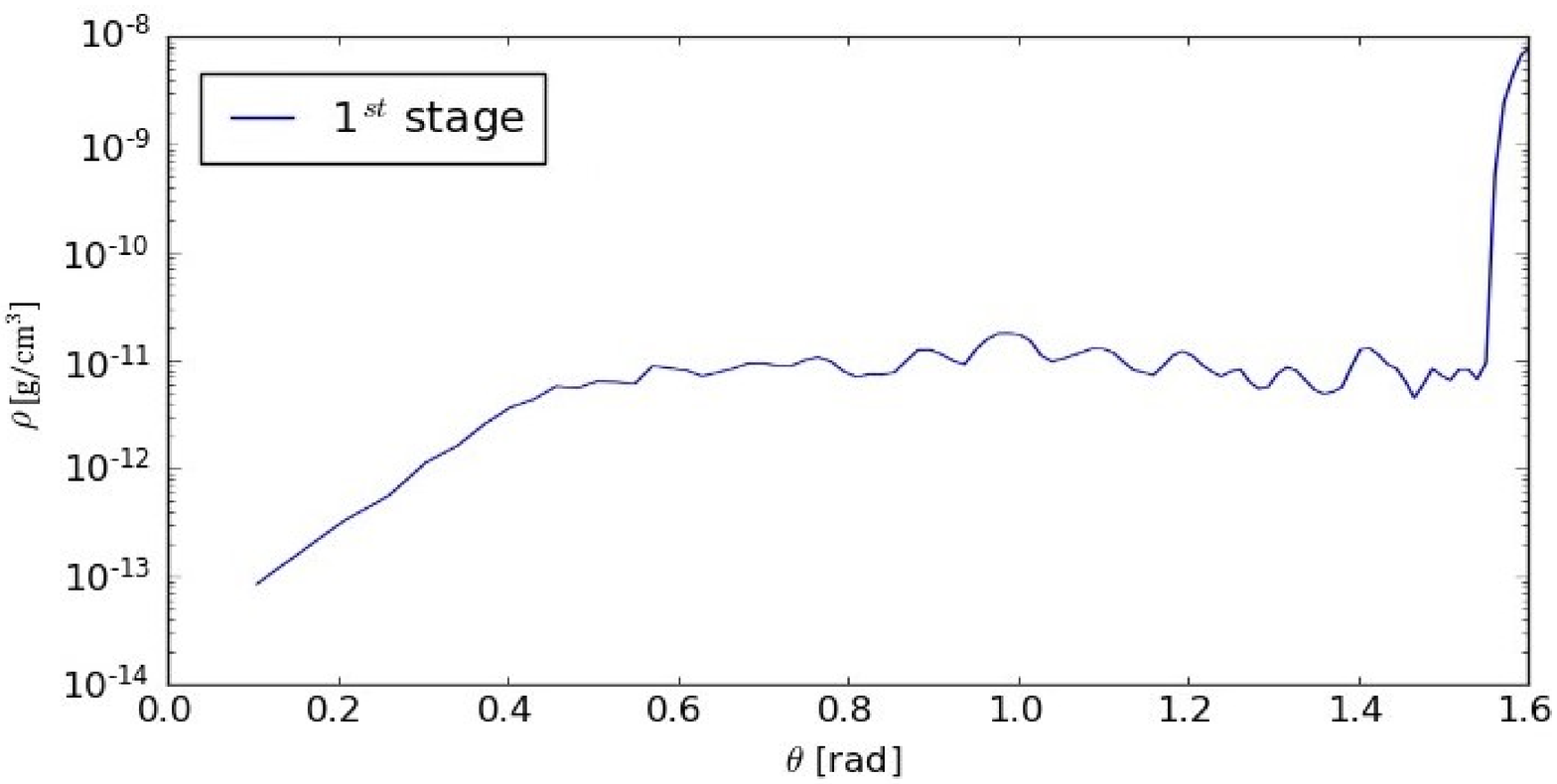}
    \includegraphics[width=0.45\textwidth]{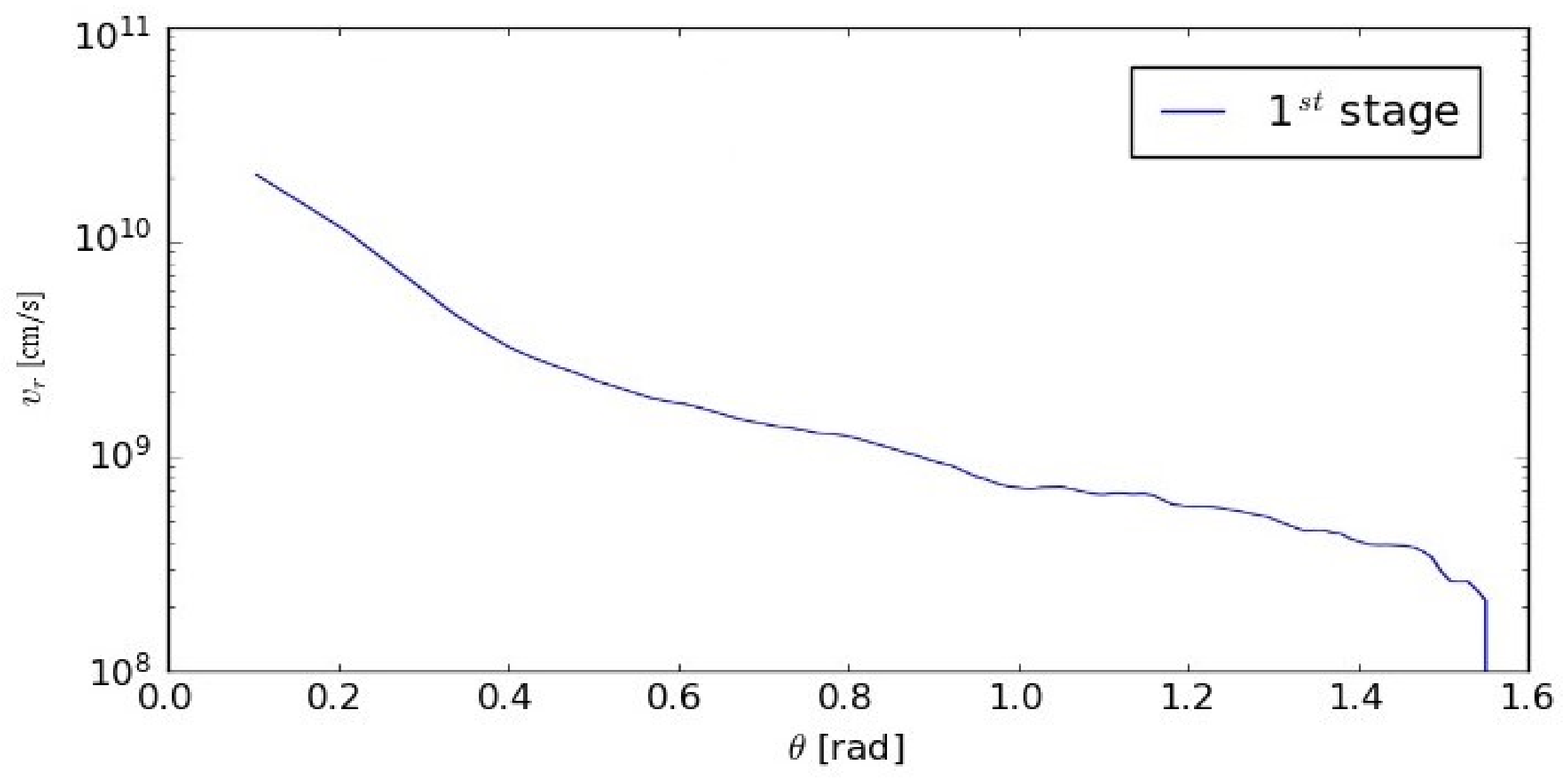}
  \end{centering}    
    \caption{Density (top) and radial velocity (bottom) angular profiles near the outer boundary of the inner zone. Notice that the radial velocity in the inflow region (i.e., $\theta\sim \pi/2$) diverges due to its negative values.}
    \label{Step_1_dens_fit}
\end{figure}

We can study the outflow properties near the outer boundary of this inner zone ($r\sim 3000r_{\rm{Sch}}$), to understand what to expect in the next stage. To do this, we look at the azimuthal profiles of hydro-dynamical variables near the outer boundary (as shown in Figure \ref{Step_1_dens_fit}), where we see 3 distinct regions:

\begin{itemize}
    \item $\theta_{\rm{ah}}<\theta \leq \pi/2$: Inflow region, high density gas with negative velocity.
    \item $0.4<\theta<\theta_{\rm{ah}}$: Uncollimated outflow region, constant density $\rho\sim10^{-11}$ g cm$^{-3}$, non-relativistic velocities ($0.01<v_r/c<0.03$).
    \item $\theta<0.4$: Collimated jet region, low density gas capable of achieving relativistic velocities ($v_r\sim0.7c$).
\end{itemize}

\subsection{Outflow in the 2$^{nd}$ stage}

Let us proceed to the 2$^{nd}$ stage simulation to see how outflow propagates further in the outer zone. Let us overview the flow structure, as well as radiation properties, calculated in the 2$^{nd}$ stage simulations in the right panels of Figure \ref{outflow_steps}. We see rather smooth distributions of quantities in the right panels. One may thus think that the flow structure and radiation properties in the outer zone could be simple extrapolations of those in the inner zone. It is not precisely the case, however, as will be shown later.

Since we only follow the escaping gas in this stage, we only simulate the outflow structure. This means that, due to the elimination of the inflow region by re-scaling the simulation box on the 2$^{nd}$ stage (see subsection \ref{subsec:connection}), all the results and conclusions drawn in this stage will cover only the region at $\theta < \theta_{\rm{ah}} \sim 1.5 \, \rm{rad}$. This value is the limit for which $v_r(\theta)<0$, as seen in the bottom panel Figure \ref{Step_1_dens_fit}. In the top panel of the same figure we can confirm that $\theta \sim 1.5 \, \rm{rad}$ delimits the inflow region from the almost 4 order magnitude drop in density.

\begin{figure}[ht!] 
  \begin{centering}
    \includegraphics[width=0.5\textwidth]{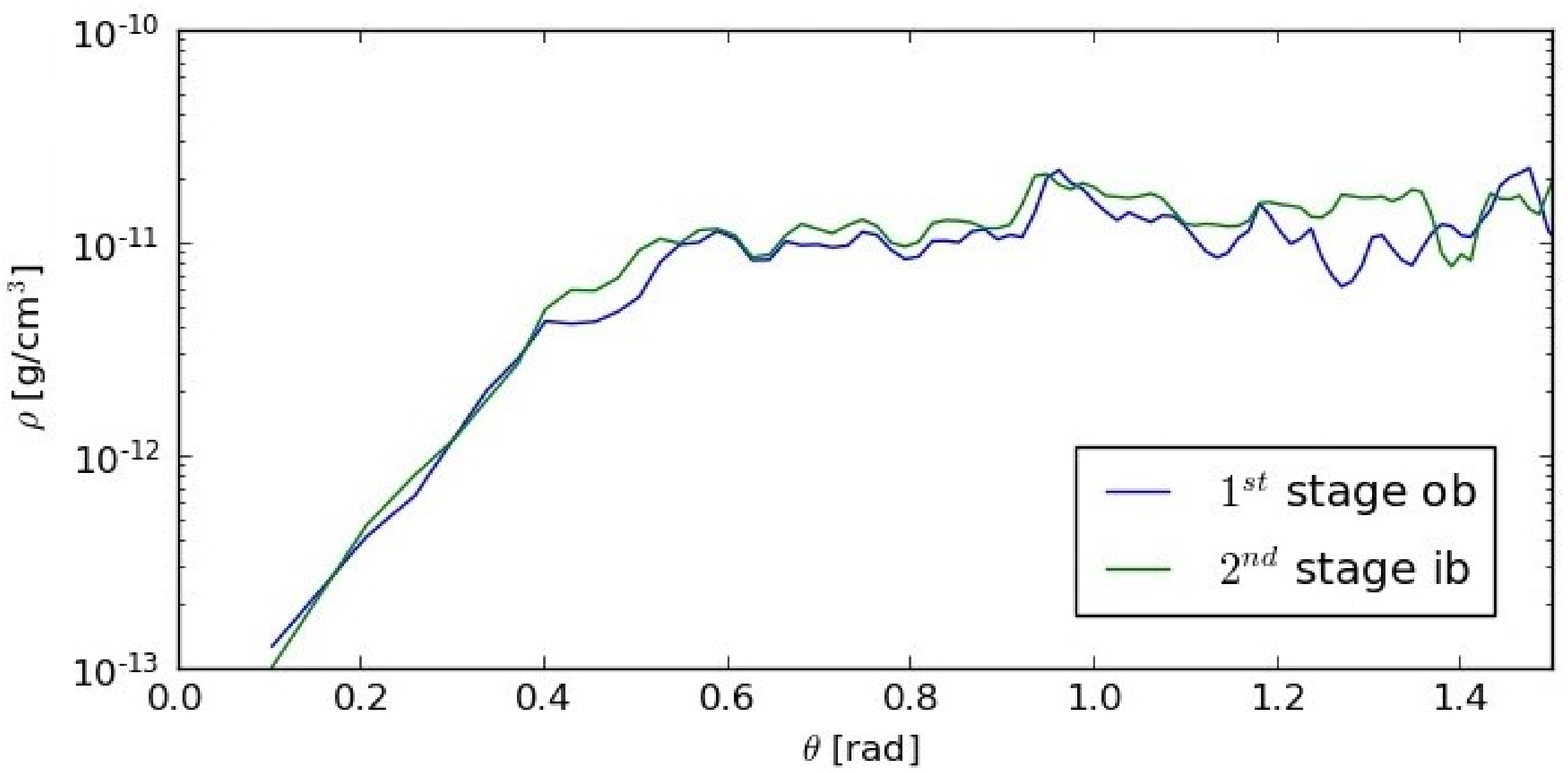}
    \includegraphics[width=0.5\textwidth]{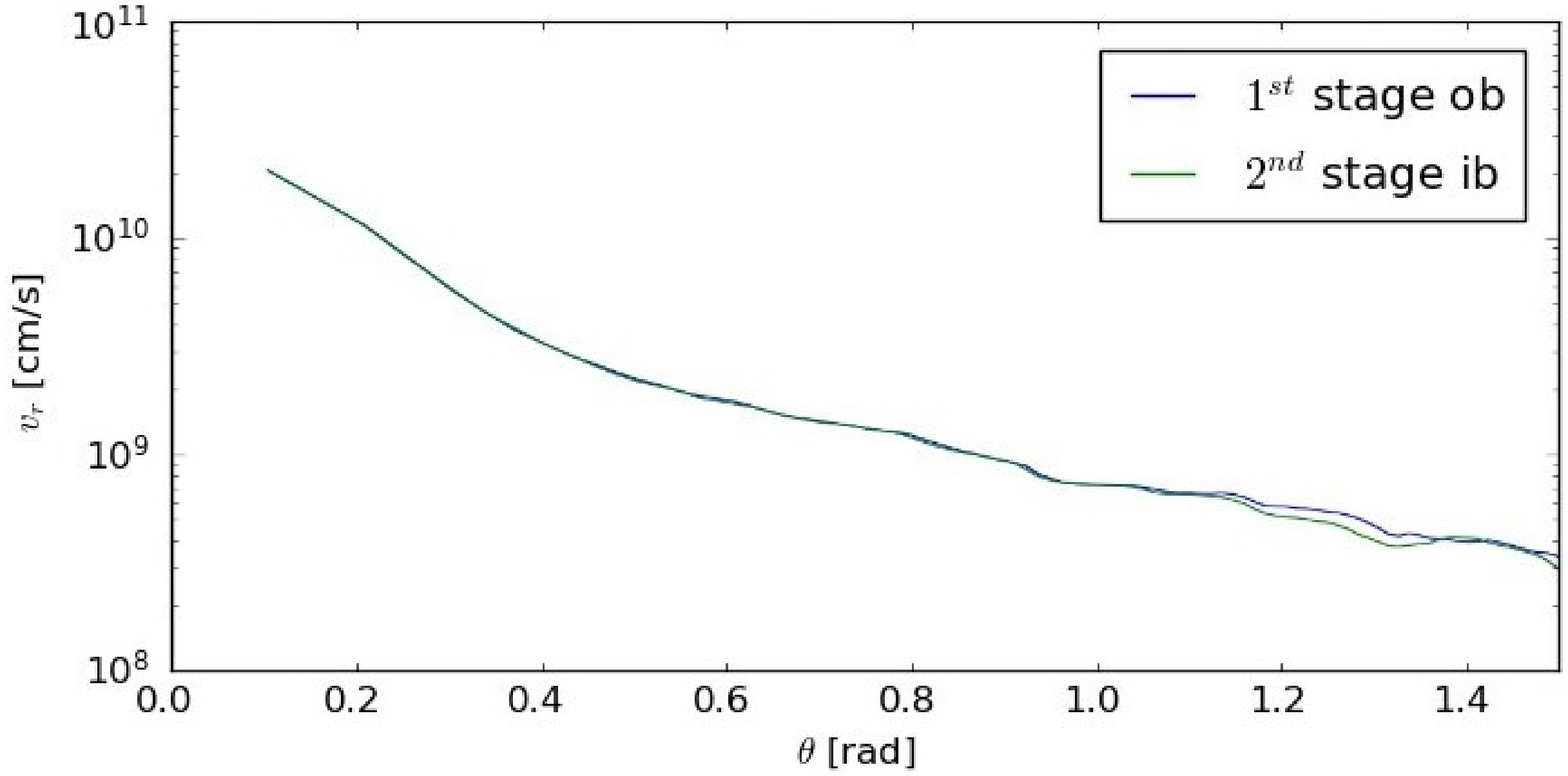} 
    \includegraphics[width=0.5\textwidth]{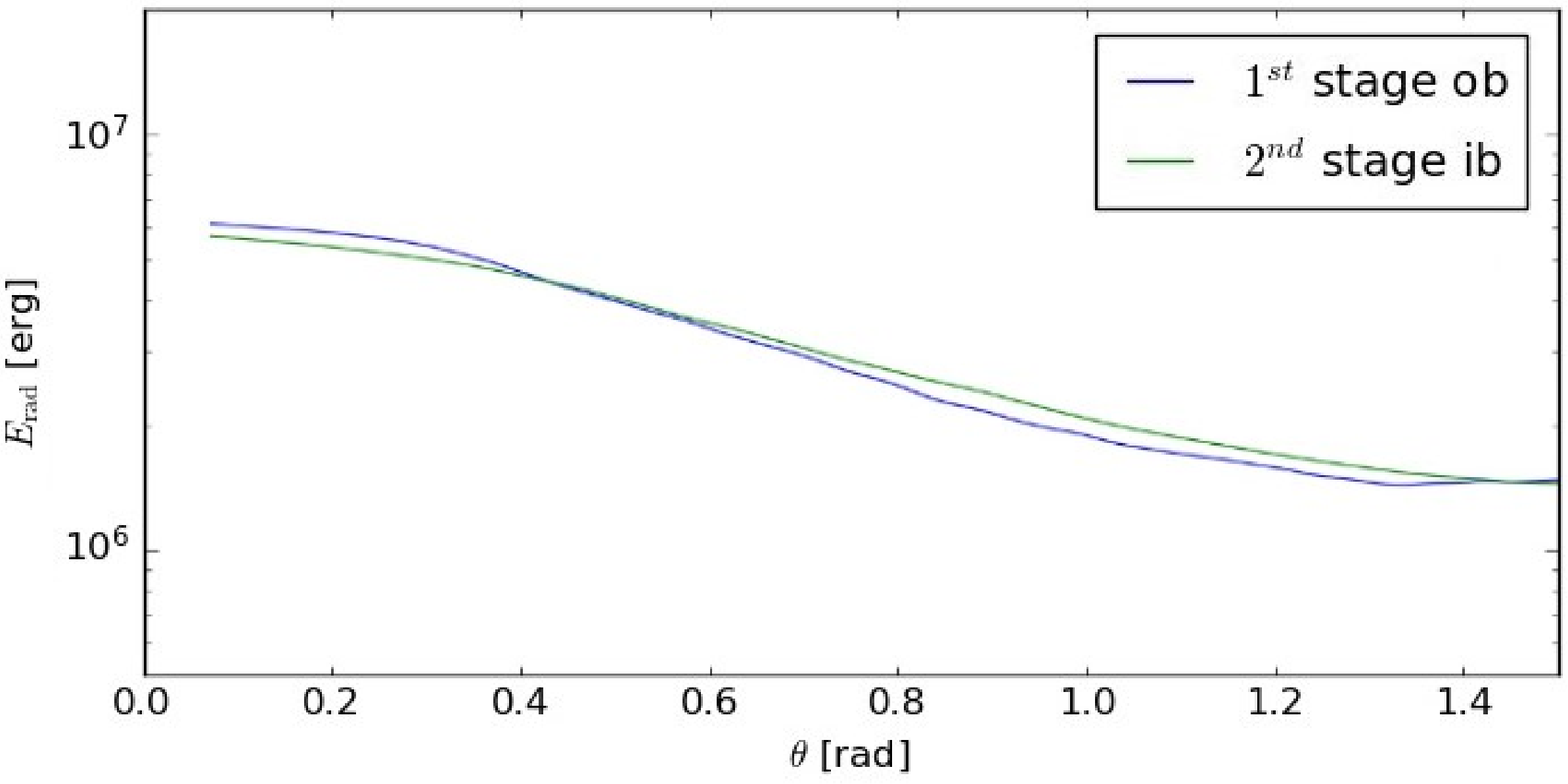} 
  \end{centering}
    \caption{Angular profiles of the density (top), the radial velocity (center) and the radiation energy (bottom) measured at the same radius $r=2.7 \times 10^3 r_{\rm{Sch}}$ in the inner zone (blue line) and the outer zone (green line), respectively. This graph showcases the degree of fidelity in which we can reproduce the previous zone result.}
    \label{2nd_vs_3rd}
\end{figure}

In section \ref{sec:method} we discussed the challenges regarding how the program must be adapted and modified for this stage to work. Given these approximations, it is important to showcase the accuracy of our method to connect 2 stages. For this reason, we show some of the physical quantities measured in both of 1$^{st}$ and 2$^{nd}$ simulation boxes at the same radius. Figure \ref{2nd_vs_3rd} shows the density profiles, along with radial velocity and radiation energy density, measured in the connection zone shared with the inner and outer zones. This ensures the smooth connections between the two stages.

\begin{figure}[ht!] 
  \begin{centering}
    \includegraphics[width=0.48\textwidth]{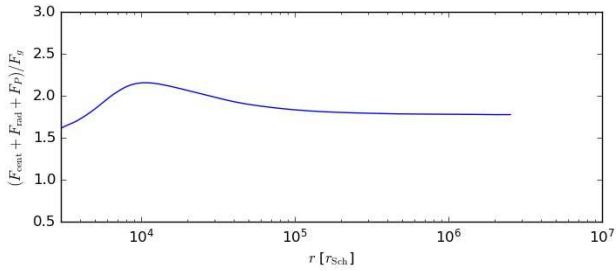} 
  \end{centering}
    \caption{Force balance between the accelerating forces ($F_{\rm{rad}}$, $F_{\rm{cent}}$ and $F_{p}$) against the gravitational pull ($F_{g}$) in the 2$^{nd}$ stage simulation, averaged over time. We have chosen $\theta\sim 0.87 \rm{rad}\sim 50^{\circ}$ as a representative value of the outflow. This figure demonstrates acceleration of the gas in the outer zone.} 
    \label{force_balance}
\end{figure}

In this stage we can also perform a study on the the mechanism of acceleration of wind along the simulation box. For this we can use equation \ref{r_motion}, which shows that the forces involved in the gas acceleration are: the radiation force (i.e., $\chi/c \, F_{0,r}$), the internal force (i.e., $\partial p/\partial r$), and the centrifugal force (i.e., $\rho (v_{\theta}^2+v_{\phi}^2)/r$). We see in Figure \ref{force_balance} that when compared with the gravitational pull, the other forces dominate. If we break down the contribution we find that the radiation force is responsible for $90-99\%$ (from $10^3-10^6 \, r_{\rm{Sch}}$) of the push and that the rest comes almost entirely from the centrifugal force. 

We have mentioned that the minimum length scales which cosmological simulations can reach is approximately $r\sim10^9 r_{\rm{Sch}}$, whereas our 2$^{nd}$ stage simulation can cover the range up to $r\sim 10^6r_{\rm{Sch}}$. Hence there is a gap between them. However, we will demonstrate that the outflow properties can well be extrapolated to even larger radii. This will be attempted in the next subsection.

\subsection{Outflow impact on the cosmological scale}

As mentioned in Section \ref{sec:intro}, it is our main objective to provide information regarding what impact the outflow can give to the environmental gas. This is because such impacts are simply assumed or modeled without justification in most cosmological simulations. For this reason, those cosmological simulations tend to miss such an important physical process. But thanks to our nested simulation-box method we can precisely evaluate the mass, momentum, and energy fluxes generated by the outflow winds.

\begin{figure}[ht!] 
  \begin{centering}
    \includegraphics[width=0.5\textwidth]{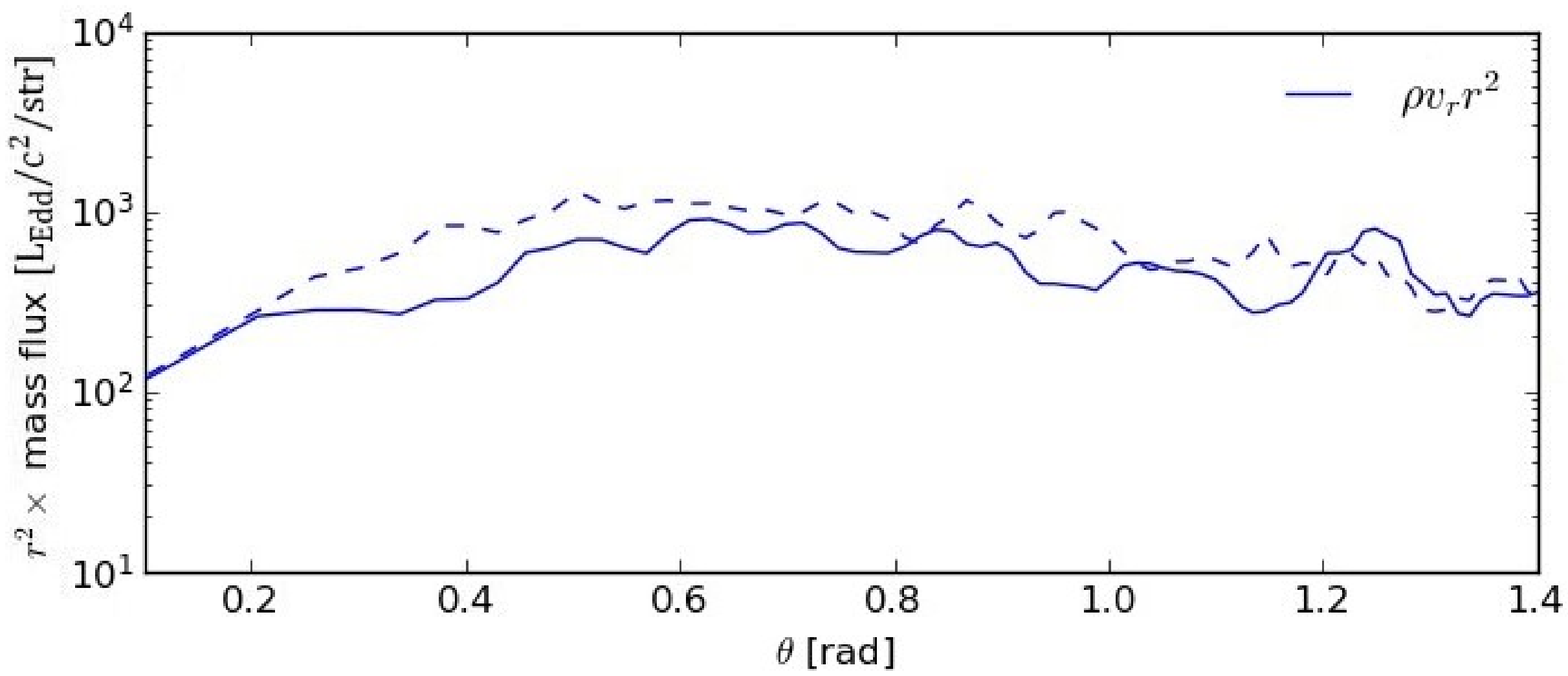}
    \includegraphics[width=0.5\textwidth]{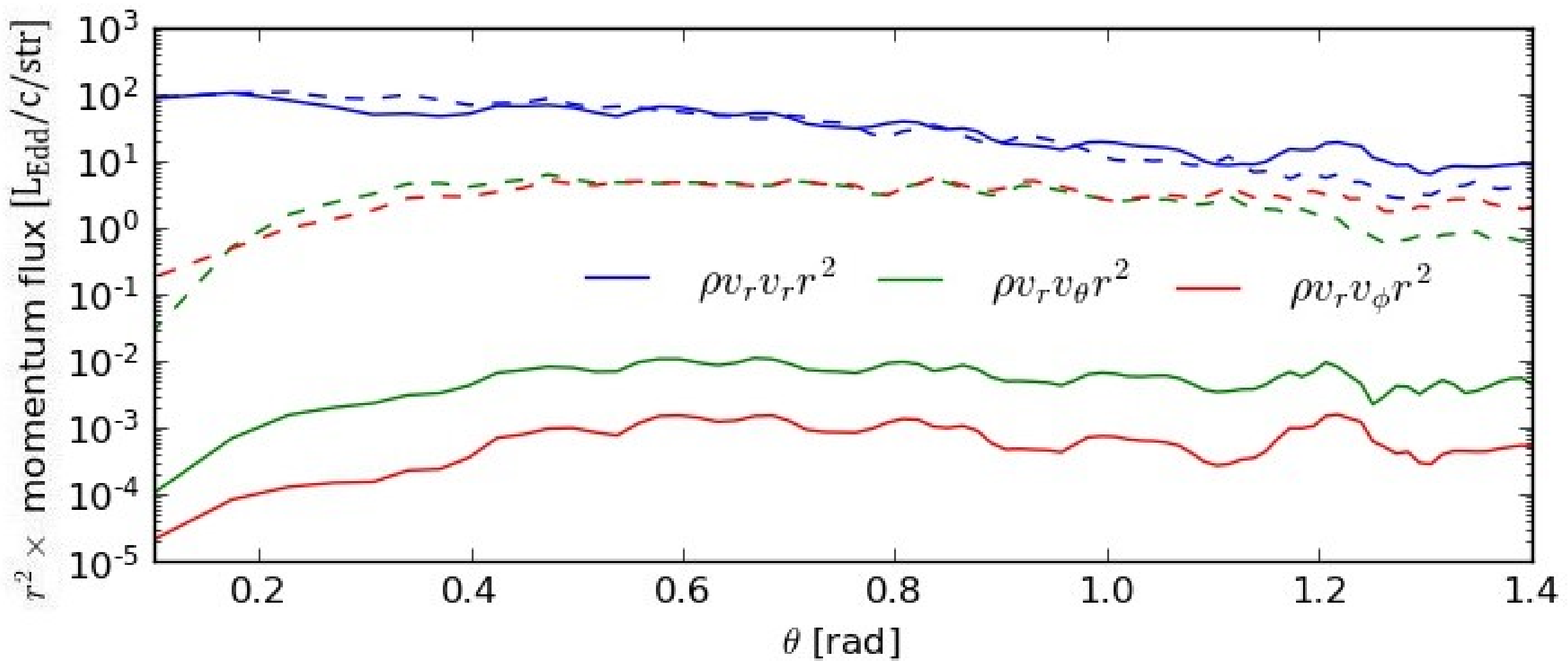}
    \includegraphics[width=0.5\textwidth]{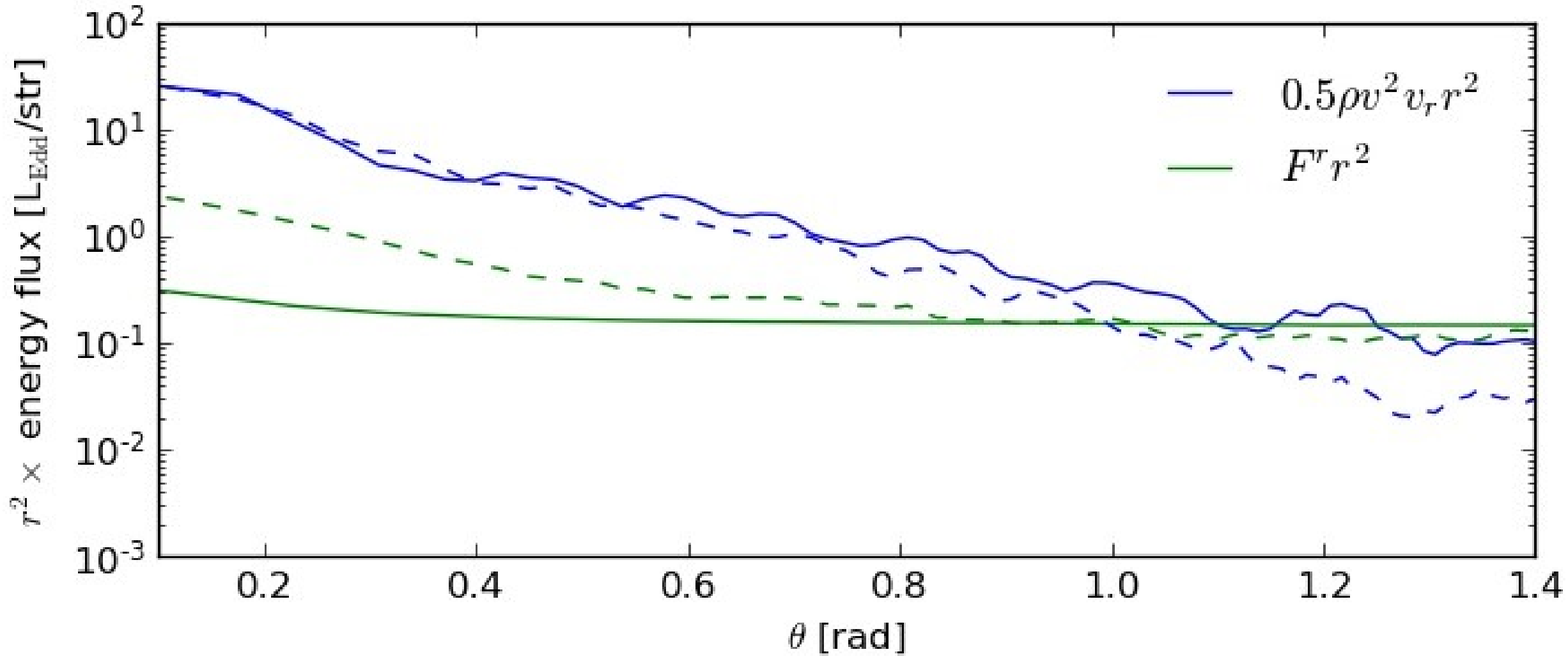}
  \end{centering}    
    \caption{Linear profiles of mass (top), momentum (center) and energy (bottom) flux in the angular direction ($0.0<\theta<1.4$). Solid lines indicate the flux at the outer boundary of the 2$^{nd}$ stage (i.e., $r\sim3 \times 10^6r_{\rm{Sch}}$), while the dashed line shows the flux at the outer boundary of the inner zone (i.e., $r\sim3\times10^3r_{\rm{Sch}}$).}
    \label{fluxes}
\end{figure}

In Figure \ref{fluxes} we show the angular profiles of the mass, momentum, and energy fluxes (all multiplied by $r^2$) at the outer boundaries of the 1$^{st}$ and 2$^{nd}$ stage simulations. Here we calculate not only the total luminosity but also the isotropic X-ray luminosity and the isotropic mechanical luminosity defined as:

\begin{equation}
    L_{\rm{X}}^{\rm{ISO}}(\theta)=4\pi r^2 F^r,
\end{equation}

\begin{equation}
    L_{\rm{mec}}^{\rm{ISO}}(\theta)=2\pi r^2 \rho v^2 v_r,
\end{equation}

\noindent where $v = \sqrt{v_r^2+v_\theta^2 + v_\phi^2}$, $F^r$ is the radiation flux in the laboratory frame and we assume that radiation is emitted predominantly in the X-ray band, since the ratio between the X-ray luminosity to the bolometric luminosity is $\sim 71-98\%$ from \citet{2017PASJ...69...92K} and \citet{2017MNRAS.469.2997N}.

In Figure \ref{fluxes} we find that the mass flux profile is nearly flat, but the momentum and energy fluxes tend to grow as the azimuthal angle $(\theta)$ decreases (towards the rotation axis). That is, we expect much larger impacts to the environments located in the face-on direction. This is because the gas mass density rapidly decreases (with decreasing $\theta$), while the radial velocity increases. These results are consistent with those of \citet{2021arXiv210111028K} (see their Fig. 14).

We have pointed through this paper the importance of moving away from the BH scale to the cosmological scale in order to paint the better picture for the outflow structure. This can be exemplified when comparing the dashed lines with the solid ones in each panel of Figure \ref{fluxes}. From the top panel, we can understand that if we would study the outflow structure solely in the inner zone (i.e., calculations only in the 1$^{st}$ stage simulation box), we would overestimate its impact by a factor of $\sim$ 2, at most. This is due to the gravitational pull of the central object; that is, the outflow with velocity being less than the escape velocity cannot reach the infinity. This is what we called failed outflow (see section \ref{subsec:1st_inflow} discussion).

In the middle panel of Figure \ref{fluxes} we can see that radial momentum flux far exceeds the other components. We also find that the $\theta$- and $\phi$- momenta decrease outward, but these can be easily understood from the viewpoints of the angular momentum conservation. In fact, they are about 3 orders of magnitude larger in the 1$^{st}$ stage (measured at $3 \times 10^3 r_{\rm{Sch}}$) than in the 2$^{nd}$ stage (measured at $3 \times 10^6 r_{\rm{Sch}}$).

In the last panel (bottom), we see how the energy impact is mostly dominated by the mechanical flux, except near the equatorial plane. We can also see how the impact from radiation flux increases towards the rotation axis (with a decrease in $\theta$). At the same time its impact is reduced as we increase the radius, and becomes almost spherically symmetric near $r\sim 3\times 10^6 r_{\rm{Sch}}$ (i.e., the radiation flux in the 2$^{nd}$ stage (see the solid green line) has a nearly flat profile and, overall, it is smaller than the same flux in the 1$^{st}$ stage (see the dashed green line)). This radial decrease in the impact caused by the radiation flux predicts a mechanical dominated energy impact at the cosmological boundary ($r\sim 10^{8-9}r_{\rm{Sch}}$).  It is important to point that the mechanical energy flux, in the outer zone, is by over one order of magnitude greater at large $\theta$ than that in the inner zone, indicating that it does increase radially as the outflow propagates outward. This is due to the continuous acceleration of outflowing gas by receiving radiation pressure force.

\begin{figure}[ht!] 
  \begin{centering}
    \includegraphics[width=0.5\textwidth]{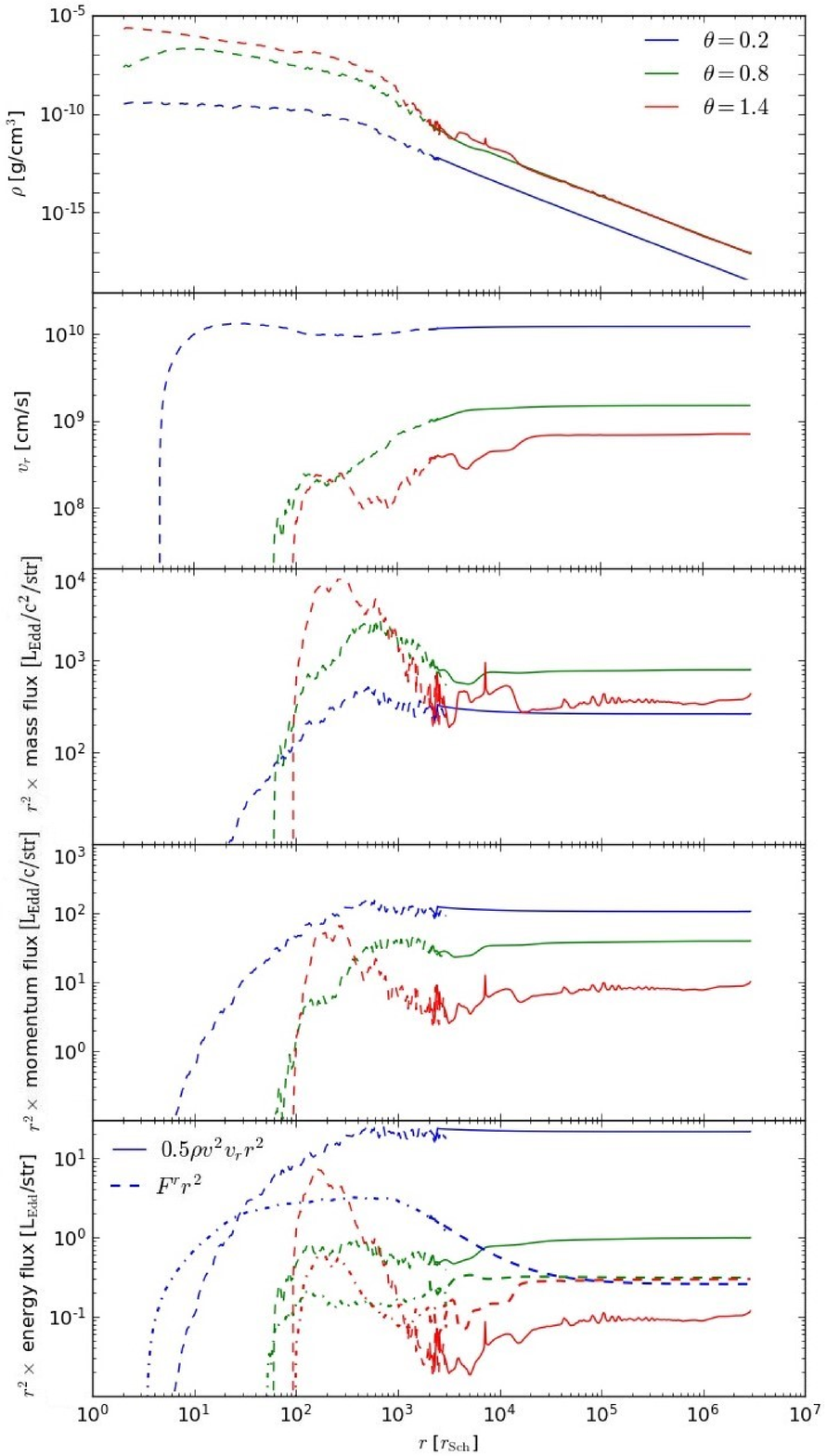}
  \end{centering}
    \caption{From top to bottom: Linear profiles of density, radial velocity, radial mass flux ($\rho v_{r}r^2$), radial momentum flux ($\rho v_{r}^2r^2$) and energy flux in the radial direction. Each panel shows both the 1$^{st}$ (dashed) and 2$^{nd}$ (solid) stage lines. The energy flux panel (last) contains also the radiation flux for the 1$^{st}$ and 2$^{nd}$ stages, indicated by a dotted and a dot-dash line respectively.}
    \label{not_extra}
\end{figure}

In order to more explicitly demonstrate the acceleration of outflowing gas and the continuous increase of mass outflow rate at small $\theta$ we show the radial profile of the gas density, radial velocity, as well as those of mass, momentum, and energy fluxes for fixed angles of
$\theta = 0.2$ ($\sim 11^{\circ}$), 0.8 ($\sim 46^{\circ}$),   
and 1.4 ($\sim 80^{\circ}$) in Figure \ref{not_extra}.
These plots not only demonstrate the smooth connection of physical quantities between the two stage simulations but also show rather uniform structure in the 2$^{nd}$ stage simulation. In this sense, the results of the 2$^{nd}$ simulation can be predicted by those of the 1st simulation, but there is one important exception. That is, the mechanical energy flux (shown in the bottom panel) shows a clear tendency of increase with an increase of radius for the nearly edge-on case (with $\theta = 1.4$). This is the direct evidence of continuous acceleration of outflow travelling in the nearly edge-on direction.

To accurately establish the impact of the outflow into the outer medium, it is also important to know  what the total mass flux is at the outermost boundary (i.e., the outer boundary of the outermost box). If we proceed in a parallel manner as Equation (\ref{Mout_eq}), we obtain that $\dot{M}_{\rm{out}}(r=3 \times 10^6r_{\rm{Sch}})\sim 317.5 L_{\rm{Edd}}/c^2\approx 0.3 \dot{M}_{\rm{inj}}(r=3 \times 10^3r_{\rm{Sch}})$. This implies that from the input material, we are losing 40\% of it as outflowing wind. 

We can use the 2$^{nd}$ stage data to extrapolate to cosmological scales ($\sim 0.1$ pc). This is because, inside the computational box in this stage, $\dot{M}_{\rm{out}}=\dot{M}_{\rm{esc}}$ is reached at $r\sim 6000r_{\rm{Sch}}$, thus the outflow measured at the outer boundary will contain no failed outflow. The radius at which this condition is met seems to depends on several factors ($\dot{M}_{\rm{inj}}$, $r_{\rm{Kep}}$, $M_{\rm{BH}}$), but we need further studies to specify which factor is most essential. For this we can use the profiles shown in Figure \ref{not_extra}, where we can see how the density and velocity (first 2 panels) trends become smooth for $r>10^5 r_{\rm{Sch}}$. In particular we see how at the larger scales $\rho\propto r^{-2}$ and $v\propto r^0$. Using these trends, we obtain that $\rho(10^9r_{\rm{Sch}})\sim10^{-23}-10^{-25}\rm{g}/\rm{cm}^3$, $v_r(10^9r_{\rm{Sch}}=3\times10^9 - 2\times 10^{10} \rm{cm}\;\rm{s}^{-1}$, for $\theta=0-1.5$. This is particularly important, since it allows us to compare our results with cosmological studies that assume a certain outflow from the central particle (\cite{2012MNRAS.420.2221D}). In their study they assume a sub-Eddington outflow defined by $v_r\sim 10^4 \rm{km}\,\rm{s}^{-1}=10^9\rm{cm}\,\rm{s}^{-1}$, momentum flux $\sim \tau_w L/c$ and energy flux $\sim 0.01 \tau_w L$, where $\tau_w=1-10$ and the luminosity $L=\min(0.1\dot{M}_{\rm{in}}c^2,L_{\rm{Edd}})$. From Figure \ref{not_extra}, by contrast, we find momentum flux $\sim 10-100 L/c$ and energy flux $\sim 0.1-10 L$ (where our luminosity is obtained from the last panel of Figure \ref{fluxes} as $L\sim0.25 L_{\rm{Edd}}$) with stronger impacts produced towards the nearly face-on direction than in the nearly edge-on direction. Our model predicts in overall stronger impacts on the environments, with a similar radial speed (i.e., $v_r\sim 10^9 \rm{cm s}^{-1}$) and energy to momentum flux relation (i.e., energy flux about 100 times smaller than momentum flux) for $0.7<\theta<1.5$, for the super-Eddington accretion scenario. The presence of a stronger wind in our model is expected from our super-Eddington scenario (which is effective when the black hole mass is relatively small) when compared to sub-Eddington scenarios, but our model can account for the presence of a collimated relativistic jet, which produces a bigger impact in the azimuthal direction. This anisotropy in the velocity is not found in wind models in other studies.

\section{Discussion} \label{sec:discussion}

In section \ref{sec:intro}, we presented the shortcomings of both cosmological and astrophysical simulations. We remarked the difficulty in solving the bridge between them due to enormous computational times. In our project the inner zone (i.e., $2r_{\rm{Sch}}<r<3000 r_{\rm{Sch}}$) took approximately 3 weeks to simulate. Increasing the radius of the outer boundary by an order of magnitude, while keeping the inner boundary position and the number of grid points constant, would make the simulation cost increase enormously. If we would increase the radius of the outer boundary, while keeping the same grid point number, we would also be losing resolution in the smaller scale. This would result in a grid size incapable of resolving correctly the circulating structure thus being incapable of correctly accounting for outflow or inflow values. The only way to overcome this would be to increase the grid point density together with the outer boundary which would increase the computational time too much, making it unfeasible. By using the nested simulation-box method, by contrast, we can save time. In fact, the 2$^{nd}$ stage (i.e., $2000 r_{\rm{Sch}}<r<3\times 10^6r_{\rm{Sch}}$) took only 5 days or so. In total the simulation cost of by using our method allowed us to reduce an impossibly long simulation to simply one month.

From Figure \ref{M_in} we find that, our model predicts a mass growth of $\dot{M}_{\rm{BH}}\sim 0.4 \dot{M}_{\rm{inj}}$. By comparing our results with those by past studies (see a compilation in Table 1 of \cite{2021arXiv210111028K}), we can gauge that, the $\dot{M}_{\rm{out}}/\dot{M}_{\rm{BH}}$ ratio is larger than in other scenarios. That implies that, for other simulation parameters (i.e., higher angular momentum, different seed size, etc.), we would expect a lower percentage of the mass being expelled by the central object, and a higher inflow rate. That being said, an efficiency of $100 \%$ mass growth, which most cosmological simulations assume, is still grossly overestimated. Thus, any conclusion on the capability of raising a super sized SMBHs is circumstantial. It is important to note that higher accretion rates still could be feasible, if RHD simulations would include some relativistic accretion mechanisms (e.g. the Blandford-Znajek process \citet{2021arXiv210210649K}).

We have also had to modify the opacity function at large scales to include the recombination of hydrogen at low temperatures. We used a simplified formulation to include this factor when temperature drops below $T<10^4 K$. This needs to be included since, as seen in \citet{2018MNRAS.476..673T}, the impact of the gas chemistry in the medium is non-negligible. 

Besides the inflow study, our method, as we showed, also allowed us to trace the jet and outflow structure of such bodies. Powerful outflow, if exists, would collide with inflow gas stream, thereby being able to suppress the gas inflow motion. Such effects were not properly considered in the cosmological simulation (e.g. \citet{{2016MNRAS.456..500S}}). In fact, these studies either do not include AGN feedback or adopt just a simplistic model. The extension of the outflow effects on such scales is dependent on the pressure exercised by the accretion structure. 

We have also found that the impacts at $\sim$ 0.1 pc are by a factor of around 10 times larger than the one assumed by \citet{2012MNRAS.420.2221D}. This difference stems from much larger velocities and highly anisotropic velocity profile (see the middle panel in Figure \ref{2nd_vs_3rd}), whereas they assumed the spherical symmetric non-relativistic velocity profile. From \citet{2012MNRAS.420.2221D} we see that momentum fluxes $\gtrsim 3 L/c$ are expected to produce AGN-driven galactic outflows capable of suppressing star formation and accretion in the host galaxy. Given the results obtained in our model, where momentum flux $10-100 L/c$, we expect a larger impact of the global interstellar gas, than that assumed in cosmological simulations. 

It is relevant to note that larger velocities in the present simulation may be due to small $r_{\rm{Kep}}=100r_{\rm{Sch}}$, which may lead to larger outflow rates (see discussion in \citet{2021arXiv210111028K}, where they assumed relatively large Keplerian radius (i.e., large angular momentum of the injected gas), and find no puffed-up structure and significantly smaller mass outflow rate).

In future, we wish to continue improving the nested simulation-box method. As previously mentioned, our method was applied to a simplified case as a framework to develop and test our methodology. Moving forward we wish to challenge the nested simulation-box method by applying it to more realistic scenarios. In particular, we wish to study cases with a more realistic initial angular momentum where the gas is not able to free-fall so close to the BH boundary. There is also the need to incorporate the general relativistic (GR) effects, as well as to analyze and develop a better physics environment for the even larger simulation box. Specifically, we wish to introduce a better approach to the energy diffusion due to b-f collisions when the gas passes the ionized temperature threshold. 

\section{Conclusions} \label{sec:conclusions}

We started this study by introducing the extreme difficulty of performing a thorough study the gas dynamics around the high redshift AGN. In particular we focused on the disconnection between the studies of these objects in BH astrophysics and cosmology. Through the implementation of the nested simulation-box method we have managed to bridge that gap. In this method we first calculate inflow-outflow structure in the inner zone and follow the outflow propagation in the outer zone with smooth connection between them. The nested simulation-box method allows us to follow the evolution of outflow structure in a self-consistent fashion from astronomical scale to cosmological scale, while maximizing the computational efficiency. What we find with this method, can be summarized as follows:

\begin{itemize}
    \item Under an inflow rate of $\dot{M}=10^3 L_{\rm{Edd}}/c^2$, the accretion rate onto the central object is $\dot{M}_{\rm{BH}}\sim 0.4 \dot{M}_{\rm{inj}}$ (see Figure \ref{M_in}). That is, about 60\% of gas is lost as outflow.
    
    \item We confirm the presence of failed outflow, which is launched at smaller radii but falls back to the disk at larger radii (top and bottom panels of Figure \ref{outflow_steps}). Due to the presence of such a large-scale circulating flow we need to take sufficiently large simulation box in the inner zone.
    
    \item Outflow structure is much simpler in the outer zone and can be extrapolated beyond the outmost boundary of our simulation ($r=10^6 r_{\rm{Sch}}$) (see Figure \ref{outflow_steps}). The possible impacts of the outflow on the cosmological scale are demonstrated in Figure \ref{fluxes}.
    
    \item Simple extrapolations of the physical quantities calculated in the inner zone do not work (see Figures \ref{fluxes} $\&$ Figures \ref{not_extra}).
    
    \item Our results indicate that much larger and highly anisotropic impacts are expected on the cosmological scale ($\sim$ 0.1 pc) than those assumed in the cosmological simulation studies (as discussed in section \ref{sec:discussion}).
    
    \item In this work we applied our method to a simplistic accretion scenario. In order to make it easier to develop, we used some approximations that will be improved in future work. Parameter studies (for various initial angular momentum and mass injection rate) is to be attempted in future work.

\end{itemize}


\begin{ack}
Numerical computations were mainly carried out on Cray XC50 and the analysis servers at Center for Computational Astrophysics, National Astronomical Observatory of Japan. This work was supported in part by JSPS KAKENHI Grant Number JP18K13594 (TK), JSPS Grant-in-Aid for Scientific Research (A) JP21H04488 (KO), same but for Scientific Research (C) JP20K04026 (SM) and JP18K03710 (KO).This work was also supported by MEXT as "Program for Promoting Researches on the Supercomputer Fugaku" (Toward a unified view of the universe: from large scale structures to planets, JPMXP1020200109) (KO and TK), and by Joint Institute for Computational Fundamental Science (JICFuS, KO). KA is financially supported in part by grants from the National Science Foundation  (AST-1440254, AST-1614868, AST-2034306). 

\end{ack}


\end{document}